# Biologists need modern data infrastructure on campus

Andrey Andreev*, Tom Morrell, Kristin Briney, Sandra Gesing, Uri Manor

Correspondence: aandreev@caltech.edu

Last updated: August 16, 2021

Modern tools for biological research, especially microscopy, have rapidly advanced in recent years [1][2], which has led to the generation of increasingly large amounts of data on a regular basis. The result is that scientists desperately need state-of-the-art technical infrastructure for raw data storage, transfer, and processing. These scientists currently rely on outdated ways to move and store data, costing valuable time and risking loss of valuable data. While the community is aware of modern approaches to data management and high-level principles (including FAIR), highly-trained and highly-paid scientists are forced to spend time dealing with technical problems, which can ultimately costs more than providing storage and a fast network on campus. Here we provide concise arguments for better infrastructure, blueprints of possible solutions, and advice in navigating the political process of solving this issue. We suggest, as a broad solution, separate NIH-managed fund for supporting universities and institutes in deployment of data storage and long-term data sharing for all funded projects.

This position statement is open for more contributors from imaging, life sciences, and other disciplines. Please contact us with your experience and perspective.

## Practice and theory of microscopy data management are disconnected

The scientific community, especially those engaged in "big data" and microscopy fields, have continually developed and refined [3][4] a strong vision for how to make data accessible, recently unified around the set of so-called FAIR [5] principles (Findability, Accessibility, Interoperability, Reuse). The scholarly communications community has developed standards for maintaining access to data over the long term, such as using persistent identifiers like DOIs, and technologies and services for efficiently transferring large amounts of data between research institutions [6]. Microscopists have developed standardized formats for images [7][8] (an extremely difficult task), better practices for metadata recording [9][10], guidelines to make data more shareable [4], impressive methods for large-scale processing and applications thereof [11], and myriad other useful data science tools. Data repositories, including cloud-based systems for Big Data, have been established [12]. Nevertheless, the practice of data management "in the trenches" remains challenging [13][14][15] and data is often unavailable even after publication [16], both in biology and other fields, for example, astronomy [17]. Almost every lab can share horror stories of lost data, failed storage servers, moving terabytes of data using consumer-oriented cloud-sharing platforms, or worse, relying on portable USB drives. Every presentation dedicated to large-data imaging experiments is followed by questions such as "how do you manage the data?". The answer too often is a sigh of disappointment and pain.

Over the years, universities and research centers such as at the Salk Institute (Waitt Advanced Biophotonics Core) and Janelia HHMI research campus (Advanced Imaging Center) have developed local data storage approaches. Nation-wide efforts have also advanced storage of scientific data, such as the German National Research Data Infrastructure or Compute Canada and the Canadian Association of Research Libraries collaboration. NSF-funded efforts such as OURRstore [18] have shown how on-premise tape libraries can be a cost-effective way of archiving large volumes of research data. Another NSF-supported project, The Open Storage network, has shown that institutions can host robust, cloud-like local storage for active data at costs substantially lower than commercial cloud providers. While funding agencies and journals are requesting better-developed plans for data management, but many scientists at large rarely have access to the resources necessary to fulfill these requests. The problem can be solved by better centralized and endowed storage infrastructure on each research campus.

## Leaky data plumbing makes data unFAIR

There are many reasons why data is being lost out of the collection-analysis-publication-archiving pipeline. Data without backups can get physically lost, data can fail to be copied, or access to data can be blocked by slow network speeds. In absence of centralized storage, a significant amount of research data is held on personal computers, or in the "cloud", inaccessible by collaborators or others in the lab. Another common problem is the required lifespan of data can be much longer than duration of the grants (less than 5 years), yet the grant is what pays for the data storage.

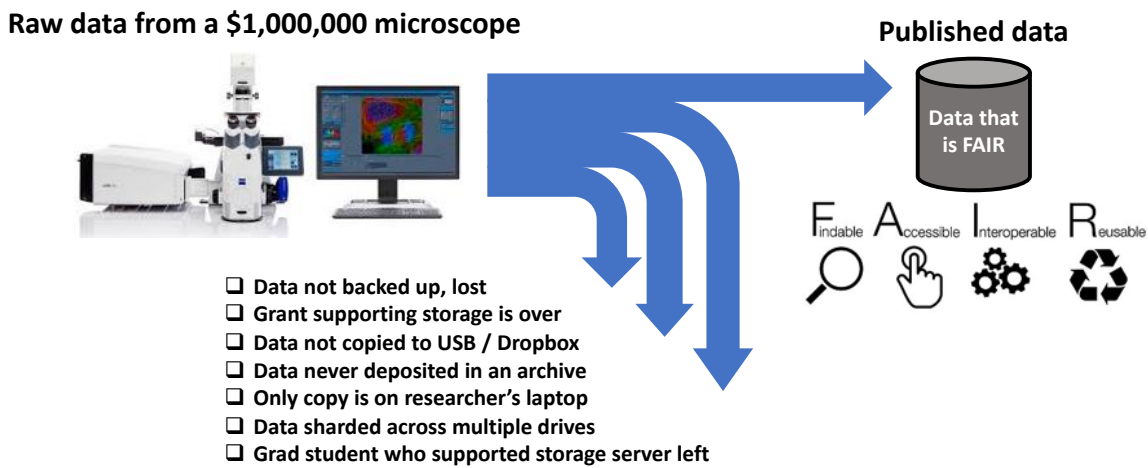

- Data not backed up, lost
- Grant supporting storage is over
- Data not copied to USB / Dropbox
- Data never deposited in an archive
- Only copy is on researcher's laptop
- Data sharded across multiple drives
- Grad student who supported storage server left

**Figure 1.** Leaky data pipelines prevent data from being FAIR

This loss of data or lack of easy access to data hurts all researchers. It hurts those who work with large datasets and small datasets, those who rely on collaborators, and those that funded the acquisition of the data, who are often on the hook for regenerating said data when it gets lost. Data issues are present on all levels of organization, from large research-centric universities to smaller institutes and colleges. The level of financial support and grant funding do not necessarily correlate with quality of infrastructure or what local IT departments offer (see Table 1 and Extended Table 1)

| University (anonymized) | Cloud storage vendors | Cloud storage limit | NAS volume, TB | NAS cost, $/TB/yr | HPC initial storage quota, TB | HPC storage cost | Research budget |
|---|---|---|---|---|---|---|---|
| 1 | Box | NA | NA | $137 | 2TB for year | $355/TB for 5yr | $800M |
| 2 | Google, Box, Microsoft | NA | 5TB per PI | $200 | ND | ND | $1200M |
| 3 | Box, Google, Microsoft | NA | NA | $460 | N/A | $1200-$3600/TB/yr | $1600M |
| 4 | Dropbox, Microsoft, Google | NA | 2GB | NA | NA | NA | $770M |
| 5 | Google, Microsoft | NA | 10TB/dept | ND | ND | ND | $330M |
| 6 | Google, Box | NA | 1TB+ | $500 | ND | ND | $800M |
| 7 | Box, Google | NA | NA | $90 | ND | ND | $1600M |
| 8 | Box | NA | NA | $560 | ND | ND | $700M |
| 9 | Microsoft | NA | 1TB+ | $120-$150 | ND | ND | $500M |
| 10 | Microsoft | NA | <90TB (first 3TB free) | $140/TB for 5 yrs | ND | ND | $130M |

**Table 1.** Survey of scientific data storage options and costs offered by various universities in the US. This table is an excerpt, information on 19 more schools is available online at [https://bit.ly/3jSoqw7]. Research budgets were estimated from corresponding Wikipedia pages. NAS: on-site offering of data storage; HPC: high-performance cluster computing storage; NA: not applicable; ND: no data can be found

This table reviews what different universities and institutes offer for data storage. Most universities offer up to three tiers of storage, that includes commercial cloud storage, limited on-campus network-attached storage (NAS), and storage on the computation cluster facilities (high performance computing, or HPC). Archiving of data using tape-based storage is a common service, but works for rarely used ("cold" storage) backups, and can be very difficult for large datasets. We only reviewed information that IT departments made available online, however a few conclusions are clear. Universities, on average, don't offer more than few TBs of NAS storage, and cloud storage options are heterogeneous, limited, and might be restricted for non-research data. The NAS and HPC costs are extremely high, and storage is often time-limited, making these impractical for long-term data storage or archiving. Thus, these options for storing scientific data are insufficient and outdated.

## Data storage is a common challenge: experience from astronomy

Scientists engaged in astronomy research face similar issues compared to biological microscopy. The amount of data keeps increasing, but reliance on data collected by others is even greater in astronomy. A detailed survey performed in 2012 [17] shows how astronomers struggle with raw and processed data storage and sharing. Self-published data (using "tilde" sites, such as personal pages) as well as curated data had a more than 20% chance of becoming unavailable within few years after publication. Astronomy community considers derived data (processed data from open sources) much more valuable than just the raw data and highlights how hard it is to share. As with microscopy, surveyed researchers stressed how data is tied to a person who shepherded or produced it (often, a graduate student). Respondents notice that data is shared using email or FTP servers, or personal websites, and only 8% of researchers used institute-based storage solutions. Cloud-based storage is a viable option, but the naïve cost is 3-7x more than local, campus-based storage [19]. Optimization can be performed to bring the cost of storage and working with data to much more reasonable numbers, but it requires that researchers have access to local infrastructure to perform that optimization.

Regardless of the challenges, data sharing is recognized as an extremely important task and even an ethical obligation in the astronomy community. After all, they study "the Universe" so it only makes sense to make all data accessible by everyone (paraphrasing one of the researchers we've talked to). Opportunity cost, that is the cost of researchers spending time dealing with hardware and data transfer, is significant and exceeds the costs of running shared infrastructure – similar to the situation in imaging-based biology research.

The astronomy community faces the same issues as biologists: the preservation of "long-lived, high-value data collections" [20] is not funded, but required. The solutions that astronomy researchers came up with include grant-driven funding of storage and sharing of data, education and outreach, and collaboration with commercial providers of IT infrastructure such as Google, Microsoft, and Amazon. Just like the biologists, astronomers require more tiers of storage on campus, not just fast and expensive storage at the high-performance computing clusters. To that end, Open Storage Network, an active NSF-funded effort to prototype storage solutions for scientific use, and specifically open-source designed petabyte-scale storage "pods" already deployed across multiple US institutes, provide a hardware solution for this storage problem.

# Centralized storage blueprint

The underlying problem of data plumbing can be alleviated by universally increasing minimum acceptable infrastructure provided by every university, institute, and research center. The state of the art in modern computer technology makes TB-scale data storage a commodity. Centralization of resources, from individual research groups to the level of the university, will drive overall costs down. We propose that the standard today must include the following features, centralized and supported by dedicated campus staff:

- On-campus, centralized, University-supported 500TB storage space per lab (or 50TB per researcher) paid through indirect costs
- Tiered storage options: 100TB of regular storage with limited lifetime (<3 years), 500TB of "cold" (slow) long-term storage for >5 years
- Automatic daily offsite backups with regular, automatic tests of backups
- Interface that allows "share-by-link" behavior to share files and folders securely with collaborators world-wide
- ≥10Gbps networks throughout the campus, especially between storage and acquisition/analysis workstations
- Public data, such as data associated with publications, should be managed by a campus organization such as the library. Such data is a critical component of the scholarly record and should be retained for decades if not indefinitely. Public data storage should be integrated into the university storage and should have features such as making data accessible at the time of publication with automated DOI assignment

We recognize that some projects need specialized solutions. In extreme cases, labs will need to take the lead in solving their unique data problems themselves. To support these needs, the organizations and IT departments, as subject matter experts, should provide templates and best practices. Following options should be universally offered:

- "Cold" storage of >100TB of selected datasets for >10 years for free
- Smaller fast storage (based on flash/SSD) solution connected via 40-100Gbps network to acquisition or analysis servers

- SSD-filled "pods" for fast transfer of very large datasets, that can be attached to acquisition systems during data collection, taken by outside users back to their home institutes, and then returned to the research institute by mail

## Responding to common arguments against centralization

Several concerns are often raised in response to calls for upgrading infrastructure. Mostly, they are financial or based on specialization. Some of the responses are provided in Table 2:

| Concern | Response |
| --- | --- |
| The university has network storage already | It is expensive or very limited (See Table 1) |
| You can use *cloud*, the university has free Dropbox license | "Cloud" as primary storage is often slow, unreliable, and limited. It is hard to share and collaborate on large datasets. It is hard to isolate equipment with sensitive data from the Internet. Some data are prohibited from deposition on any cloud storage. Commercial platforms might gain rights to use data. |
| Centralized storage is too expensive | It can cost as little as $10-30/TB/yr to store and backup data, similar to what labs pay already independently. The university doesn't need to go "enterprise" to store 1000 TB. |
| The university will need to hire IT staff | IT professionals estimate <1 hr/month support time after initial install, and storage can run on widely-used hardware and software (e.g. Dell hardware & Windows Server; no special training required). |
| Each lab should just buy its own server if you need it so much | Many labs do that right now, but it is more expensive, in money and time, and even labs with few people or resources deserve good data infrastructure. Integrating secure homebrew solutions with existing IT infrastructure is challenging |
| Everyone is used to the *status quo* | It is outdated. Newly-hired professors in imaging-heavy disciplines might need to store TBs of data, quickly. A new postdoc might come with a data-heavy project. Good infrastructure facilitates research. |
| You can just use USB drives / sticks | USB drives often fail and lose data, can be lost, and can transfer malware. Data is hard to track and organize on USBs. |
| Only ~~light-sheet~~ fancy microscopes need that | Everyone will benefit from fast, accessible, reliable storage with automatic backups. It allows pain-free data sharing, scaling up of experiments, or starting new projects. |

| People will deposit tons of their trash data | Today storage is cheap. It is more expensive to lose or create inaccessible data. Library professionals are great at curation and archiving, and will help with retiring data. Curation will allow decrease of storage needs and potential high power consumption |

**Table 2.** Addressing common concerns with providing centralized storage

## Implementation notes

When researchers raise concerns about networks and storage, they often hear "cloud" in response. However, the main function of "cloud" solutions are computation scalability and concurrency: one can access data at constant rate, no matter the number of clients. Here we take a moment to offer comparison of cost to store data. Note that both storage and moving data to and from "cloud" is expensive. The cost to keep 1TB of data for 1 year (expecting five years of life for a dedicated server) is as follows:

| Storage solution | Cost, $USD/TB/year |
|---|---|
| **Amazon Web Services (AWS) S3** | >$280 |
| **AWS Glacier** | $48 |
| **AWS Glacier Deep Archive** | $12 |
| **Dedicated *in-lab* server** | $10-30 |
| **Custom centralized campus-wide servers *or* Open Storage Network pods** | $30 |

**Table 3.** Comparison of common ways to store data in *cloud* and in lab or on campus

Not only is "cloud" storage more expensive, but there are additional issues that make it less flexible. "Cloud" storage requires more time to move data, data download costs extra, it is hard to connect to fast networks on campus, and cloud storage entails extra layers of management and administration compared to in-lab servers. As a NASA internal review highlights [21], currently "cloud" storage might not be well-suited for >200PB amount of data storage and access. Notice that the cost of dedicated servers or centralized NAS don't include the costs of real estate, labor, and hours of support (accounted for in the price of "cloud" service), which for an in-lab storage server (often supported by a graduate student) can be significant. By comparison, campus-wide solutions, built on existing infrastructure, and supported by existing IT staff can be scaled up without incurring significant costs.

If researchers decide to propose a plan for centralized storage infrastructure, there are a couple of strategies that might help. First, researchers will have to gather support among their peers (faculty and senior-level researchers) as these are the people with the most political power. If possible, researchers should try reaching out to other data-hungry department or groups, such as neuroscience, physics, or computer science. In the case where management is not supportive, there might be a chance for smaller-scale shared solutions between several labs, that can be used as an argument for larger improvements. Infrastructure requests are best timed with major building projects, such as construction of a new research facility. However, one of the authors had success with persuading their university to add 10Gbps fiber to a 12-year-old building. It is never too late to plead a case for updating old infrastructure.

Finally, always request the most you can imagine, and then some. First, it's a good idea to over-provision, and secondly, it is a better negotiating position. Most importantly, it is much easier to underestimate than overestimate future data storage needs. Data-heavy applications should be at the forefront of the conversation (such as high-content or light-sheet microscopy, high-resolution genomics, and brain activity imaging projects), driving the push for larger and faster storage. Smaller applications might be disregarded by the institution's management but will benefit from improved infrastructure nevertheless.

## We will upgrade infrastructure eventually, but let's start now – let's make an Open Storage Network for Biomedical Imaging

The history of technological progress teaches us that we will definitely arrive at a time when every lab has robust and fast multi-TB-scale data storage, but we need it to happen sooner. Providing an endowed centralized data storage is already a necessity. NIH is the perfect organization to support establishment of a separate fund that will provide data storage during funded projects, as well as archiving of valuable artifacts of the funded research. This will be directly in line with recent requirements [22] by the NIH to make funded data open and available. More ambitious goal will be larger fund for generalized, infrastructural change of data storage not only for ongoing grants. Improved infrastructure will speed up the start of new projects and collaborations, help researchers to make data available immediately before the publication, and make it harder to hide data from reviewers. Centralization of data would allow tracking the quality of data availability using automatic tools such as PresQT [23]. Having upgraded data plumbing will make it easier to work with data remotely, decrease "scientific waste", make backups more manageable. Finally, it has been already shown [24] that making raw data available and more manageable will lower the cost of scientific discovery. As research becomes more and more data-intensive, our efforts to organize and secure this data must be at the forefront of the conversation if we aim to effectively leverage the increasing amounts of data being generated. Finally, with the accelerating development of AI-based solutions, the increased accessibility and shareability of data is paramount for development of new approaches and verification of published results.

**We would like to thank everyone who helped with shaping this statement:**

- Damian Dalle Nogare (NIH)
- Valerie Thomas, Dan Koo, Francesco Cutrale, Jeremy Wiemer (USC)
- Jacqueline D Campbell, Sarah Nusser (Iowa State)
- James Jonkman (University Health Network)
- Eric Wait, Blair Rossetti (Janelia)
- Ben Steventon (Cambridge)
- Daniel Waiger (Hebrew U of Jerusalem)
- Vikas Trivedi (EMBL)
- Henry Neeman (U of Oklahoma)
- Mario Juric (U of Wash)
- Alexander Szalay (JHU)